\documentclass[onecolumn,showpacs,amsmath,amssymb,aps,pra,floatfix,nofootinbib,10pt]{revtex4-2}
\usepackage[latin1]{inputenc}
\usepackage[english]{babel}
\usepackage[T1]{fontenc}
\usepackage{mathrsfs}
\usepackage{hyperref}
\usepackage{multirow}
\usepackage{bm,epsfig}
\usepackage{graphicx}
\usepackage{subeqnarray}
\usepackage{bbold}
\usepackage{upgreek}
\hypersetup{colorlinks=true, urlcolor=blue, citecolor=blue, filecolor=blue, linkcolor=blue}
\usepackage{dcolumn}
\usepackage{amsmath}
\usepackage{color}
\usepackage{siunitx}
\usepackage{physics}

\usepackage{xcolor}

\renewcommand{\Re}{\operatorname{Re}}

\def\beq#1{\begin{equation}\label{#1}}
\def\eeq{\end{equation}}

\begin{document}

\title{%
Role of blue-shift length in macroscopic properties of high-harmonic generation
}

\author{
Margarita Khokhlova$^{1,\ast}$ and Vasily Strelkov$^{2}$
}
\affiliation{
\mbox{$^{1}$King's College London, Strand, London WC2R 2LS, UK} \\
\mbox{$^{2}$Prokhorov General Physics Institute of the Russian Academy of Sciences, 
Moscow 119991, Russia} \\
$^{\ast}$margarita.khokhlova@kcl.ac.uk
}

\begin{abstract}%
The production of brighter coherent XUV radiation by intense laser pulses through the process of high-harmonic generation~(HHG) is a central challenge in contemporary nonlinear optics. We study the generation and spatial propagation of high harmonics analytically and via {\it ab initio} simulations. We focus on the length scales defining the growth of the harmonic signal with propagation distance and show that the well-known coherence length limits HHG only for relatively low driving intensities. For higher intensities, the photoionisation of the medium, naturally accompanying HHG, leads to essentially transient phase matching and laser frequency blue shift. By systematically taking both of these factors into account, we demonstrate that the behaviour of the harmonic signal at higher intensities is defined by another length scale~--- the blue-shift length. In this generation regime the XUV intensity at a given frequency first grows quadratically and then saturates passing the blue-shift length, but the total harmonic efficiency continues growing linearly due to the linear increase of the harmonic line bandwidth. The changeover to this generation regime takes place for all harmonic orders roughly simultaneously. The rate of the efficiency growth is maximal if the static dispersion is compensated by photoelectrons near the centre of the laser pulse. Our theory offers a robust way to choose the generation conditions that optimise the growth of the harmonic signal with propagation.
\end{abstract}

\maketitle
\noindent

\section{Introduction}
Ultrashort XUV pulses have proven to be a powerful tool to explore and control the electronic dynamics in atoms, molecules and solids. This makes the table-top generation of brighter pulses an important hurdle to jump on the way to wider applications. Thus, the optimisation of the high-harmonic generation~(HHG) process, as the main workhorse used for the table-top production of ultrashort pulses, is of significant importance. One way to increase the brightness of the HHG emission is to take advantage of the microscopic properties, wisely choosing the type of the generating medium, or to advance the macroscopic properties of HHG, improving the propagation through the generating medium.

Although the problem of the phase matching for HHG was addressed already in the first studies of the process~\cite{Balcou1993, Salieres1995, Balcou1997,Constant,Rundquist1998}, it still remains actively investigated~\cite{Weissenbilder2022, Minneker2023, Finke2022, Boyero-Garci2021, Hareli2020, Fu2022}. The importance of this problem originates, in general, from the fact that even small phase advance of the driving laser field, being multiplied by a high number of the harmonic, results in non-negligible phase mismatch. 

The well-known length scales describing propagation properties of HHG are the harmonic absorption length $L_\mathrm{abs}^{(q)}$ ($q$ is the harmonic order) and the harmonic coherence length
\beq{L_coh_general}
L_\mathrm{coh}^{(q)}= \frac{\pi}{|\Delta k_q|} \, ,    
\eeq
where ${\Delta k_q} = k_{q}-q k_0$ is the detuning from the phase matching, $k_{q}$ and $k_0$ are the harmonic and the driving field wavevectors, correspondingly. The role these two spatial scales play in the optimisation of the macroscopic HHG signal was systematically studied in~\cite{Constant}. The detuning from the phase matching comes from contributions of the geometrical and medium dispersions, as well as from the dependence of the harmonic phase on the laser intensity. The former and the latter ones do not vary in time and when they dominate (typically, for early HHG experiments which used tight laser beam focusing~\cite{Sarukura1991, Sakai1994}) the assumption of the static phase matching applies. For more recent experiments a loose laser beam focusing is more common (see, for instance,~\cite{Takahashi2003, Kovacs2019, Hoflund2021}), which allows to reduce these contributions. This makes the contribution from the medium dispersion dominating. Taking into account that for XUV the real part of the refractive index differs negligibly from unity we have
\beq{L_coh}
L_\mathrm{coh}^{(q)}= \frac{\lambda_q}{2 \Delta n} \; \; \; \textrm{or} \; \; \; L_\mathrm{coh}^{(q)}= \frac{\lambda_q}{2 |1-\bar{n}|} \, ,
\eeq
where $\lambda_q$ is the $q^\mathrm{th}$ harmonic wavelength, $\bar n$ is the refractive index\footnote{Below we assume that this is the medium refractive index. Note, however, that one can obviously include in this index also contributions from the geometrical dispersion and from the dependence of the harmonic phase on the driving intensity.} at the laser frequency, and $ \Delta n = |1 -\bar n|$. 

However, the photoionisation of the medium unavoidably accompanies HHG and typically provides non-negligible contribution to $\Delta n$. This makes phase matching essentially transient. In this case $L_\mathrm{coh}$ does not give a reasonable length scale describing the phase matching because it varies during the laser pulse and even can be infinite at certain time instants. Moreover, the medium photoionisation leads to the blue shift of the laser frequency. Although small, being multiplied by the (high) order of the harmonic, it results in the pronounced blue shift of the harmonic line. (Here one can see natural parallel with the {\it spatial} variation of the phase limiting the generation phase matching). When this blue shift exceeds the initial bandwidth of the harmonic, further propagation leads to XUV generation in the spectral range above this bandwidth, so this XUV does not add coherently to the initially generated one. On the one hand, it eliminates constructive interference of these fields and quadratic growths of the XUV intensity; on the other hand, it eliminates their destructive interference as well. Thus, the spectral XUV intensity saturates with the propagation, but the XUV energy grows linearly due the increase of the harmonic linewidth. The propagation distance where this mechanism turns on is the blue-shift length, which we introduced in Ref.~\cite{KhokhlovaStrelkov} as
\beq{L_bs}
L_\mathrm{bs}^{(q)} = \frac{\pi c}{q \omega_0 |n_\mathrm{f}-n_\mathrm{i}|} \; \; \; \textrm{or} \; \; \; L_\mathrm{bs}^{(q)}= \frac{\lambda_q}{2 |n_\mathrm{f}-n_\mathrm{i}|} \, ,
\eeq
where $\omega_0$ is the driving frequency, $n_\mathrm{i}$ and $n_\mathrm{f}$ are refractive indices in the beginning and in the end of the generation, and $c$ is the speed of light. 
In this paper we present the detailed analysis of the role of this length scale on the propagation properties of HHG.

\section{Theory}
\subsection{Harmonic energy \textit{vs} propagation distance} 
We start with studying the dependence of total harmonic energy on the propagation length $L$. We assume the lineshape of the microscopic harmonic response to be Gaussian, and before the propagation it is
\begin{equation}
\mathcal{F} (\omega)=\mathcal{F}_0 \exp\{ -(\omega- q \omega_0)^2/ \Delta \omega^2 \}  \, ,
\label{Gauss}
\end{equation}
where the harmonic pulse bandwidth $\Delta \omega$ (which is complex for a chirped pulse) is defined by the (real) pulse duration $\tau$ (defined as FWHM of the field intensity) as
\begin{equation}
\Delta \omega = \kappa / \tau \, ,
\label{kappa}
\end{equation}
where for a chirped pulse $\kappa$ is complex. For the bandwidth-limited pulse\footnote{There is a typo in the definition of this parameter in~\cite{KhokhlovaStrelkov}.} it is real $\kappa=\kappa_0$, where
\begin{equation}
\kappa_0 = 2\sqrt{ 2 \ln(2)} .   
\label{kappa0}
\end{equation}

As was shown in~\cite{KhokhlovaStrelkov} [see Eqs.~(13) and (14)] the harmonic intensity $I_q(L, \omega) \equiv \left | \tilde{\mathcal{E}}_q (L, \omega) \right |^2$ is written as
\begin{equation}
I_q(L, \omega) = \overline{I_q}
\left|
\mathrm{erf} \left( \gamma -\frac{\pi}{\kappa}\frac{L}{L_\mathrm{bs}^{(q)}} - i \frac{\kappa}{2}\frac{L_\mathrm{bs}^{(q)}}{L_\mathrm{coh}^{(q)}} \right) 
-
\mathrm{erf} \left( \gamma - i \frac{\kappa}{2}\frac{L_\mathrm{bs}^{(q)}}{L_\mathrm{coh}^{(q)}} \right) 
\right|^2 \, ,
\label{int_theory}
\end{equation}
where intensity $\overline{I_q}= \pi (q\omega_0 \mathcal{F}_0 \kappa L_\mathrm{bs}^{(q)}/ c)^2 e^{-(\kappa  L_\mathrm{bs}^{(q)} / L_\mathrm{coh}^{(q)})^2/2}$ does not depend on $L$, $\gamma=(\omega-q\omega_0)/\Delta \omega$ is the detuning from the exact harmonic frequency. Note that $L_\mathrm{coh}^{(q)}$ is calculated via Eq.~\eqref{L_coh}, where $\bar n$ is the refractive index in the {\it centre} of the pulse. Below we skip indices $(q)$ for the coherence and blue-shift lengths, where they are not essential. 

After some efforts the total energy of the harmonic $W_q (L)=\int d \omega I_q(L, \omega)$
can be obtained analytically from Eqs.~(13) and (14) in Ref.~\cite{KhokhlovaStrelkov} as
\begin{equation}
\begin{split}
W_q (L) = \sqrt{\frac{32}{\pi}} \Delta \omega \overline{I_q} 
\left\{
    \sqrt{\frac{\pi}{2}} 
    \Re{ 
        \left[ 
            \left( 
                \frac{\pi L}{\kappa_0 L_\mathrm{bs}} - i \frac{\kappa_0 L_\mathrm{bs}}{ L_\mathrm{coh}} 
            \right)
        \erf{
            \left(
                \frac{\pi L}{\sqrt{2} \kappa_0 L_\mathrm{bs}} - i \frac{\kappa_0 L_\mathrm{bs}}{\sqrt{2} L_\mathrm{coh}} 
            \right)
        }
       \right]
   } 
   + \sqrt{\frac{\pi}{2}} \frac{\kappa_0 L_\mathrm{bs}}{ L_\mathrm{coh}} \ 
   \mathrm{erfi} {\left( 
   \frac{\kappa_0 L_\mathrm{bs}}{\sqrt{2} L_\mathrm{coh}}
   \right)} \right. \\ \left.
   + e^{\frac{\kappa_0^2 L^2_\mathrm{bs}}{2 L^2_\mathrm{coh}}} 
   \left[
        \cos{
            \left( \frac{\pi L}{L_\mathrm{coh}} \right)
            } 
        e^{-\frac{\pi^2 L^2}{2 \kappa_0^2 L^2_\mathrm{bs}}} -1
    \right]    
\right\} \, .
\end{split}
\label{tot_energy}
\end{equation}
Note that the result for the chirped and the non-chirped pulse is the same (thus containing only $\kappa_0$, not $\kappa$) due to the analyticity of the Eq.~\eqref{tot_energy}. 

For different $L$, $L_\mathrm{coh}$, $L_\mathrm{bs}$ Eq.~\eqref{tot_energy} can be simplified as described in Appendix, see Table~\ref{tab:harm_energy_limits} for the summary.
\begin{table}[h!]
\newlength{\equationlengthtableone}
\setlength{\equationlengthtableone}{9.5cm}
\renewcommand\thetable{1}
\centering
\begin{tabular}{| c | c | c |}
\hline
Regime & Condition & Harmonic energy \\ 
\hline
&&\\
Short propagation distances
& for
\begin{minipage}{3.5cm}
\begin{equation}
 L\ll \min \left\{L_\mathrm{coh}, L_\mathrm{bs} \right\}
\label{condition0}   
\end{equation} 
\end{minipage}
&
\begin{minipage}{\equationlengthtableone}
\begin{equation}
W_q \approx 
A \frac{\pi^2}{2} \left( \frac{L}{\lambda_q}\right)^2
\label{main_tot_energy_short_L1}
\end{equation} 
\end{minipage}
\\
&&\\

Phase-matching defined
&
for all $L$ under $L_\mathrm{coh} \ll L_\mathrm{bs}$ 
& 
\begin{minipage}{\equationlengthtableone}
\begin{equation}
W_q \approx A  \left( \frac{L_\mathrm{coh}}{\lambda_q}\right)^2
\left\{ 1 - \cos\left(\frac{\pi L}{L_\mathrm{coh}}\right) \exp\left( - \frac{\pi^2 L^2}{16\ln(2) L_\mathrm{bs}^2}\right) \right\} 
\label{main_tot_energy_long_L2}
\end{equation}
\end{minipage}
\\ 
&&\\

Blue-shift defined
&
for all $L$ under $L_\mathrm{bs} \lesssim L_\mathrm{coh}$ 
&
\begin{minipage}{\equationlengthtableone}
\begin{equation}
W_q \approx  2 \sqrt{\ln(2)} \pi^{3/2} A\left( \frac{L_\mathrm{bs}}{\lambda_q}\right) \left( \frac{L+D}{\lambda_q}\right) \exp{-\frac{4 \ln(2) L_\mathrm{bs}^2}{L_\mathrm{coh}^2}}
\label{main_tot_energy_long_L5}
\end{equation}
\end{minipage}
\\
&&\\
\hline
\end{tabular}
\caption{Harmonic energy in different generation regimes.}
\label{tab:harm_energy_limits}
\end{table}
\\In the table $A$ and $D$ are defined as
\begin{equation}
A=\sqrt{32\pi} \Delta \omega (2 \pi \mathcal{F}_0)^2 \, ,
\label{A}
\end{equation}
\begin{equation}
D=\frac{4 \sqrt{\ln(2)} L_\mathrm{bs} [2 b F(b)-1] \exp{b^2}}{\pi^{3/2}} \, ,
\label{main_D}
\end{equation}
where the Dawson function $F(b) \equiv \sqrt{\pi} \exp{-b^2} \mathrm{erfi}( b)  /2$, and 
\begin{equation}
b= \frac{2\sqrt{ \ln(2)} L_\mathrm{bs}}{L_\mathrm{coh}}. 
    \label{main_b}
\end{equation}
Moreover, for $b < 1 $ Eq.~(\ref{main_D}) can be simplified as
\begin{equation}
D \approx  \frac{4 \sqrt{\ln(2)} L_\mathrm{bs} [b^2-1]} {\pi^{3/2}} \, .
\label{main_D1}
\end{equation}

\begin{figure}
\centering
\includegraphics[width=0.48\linewidth]{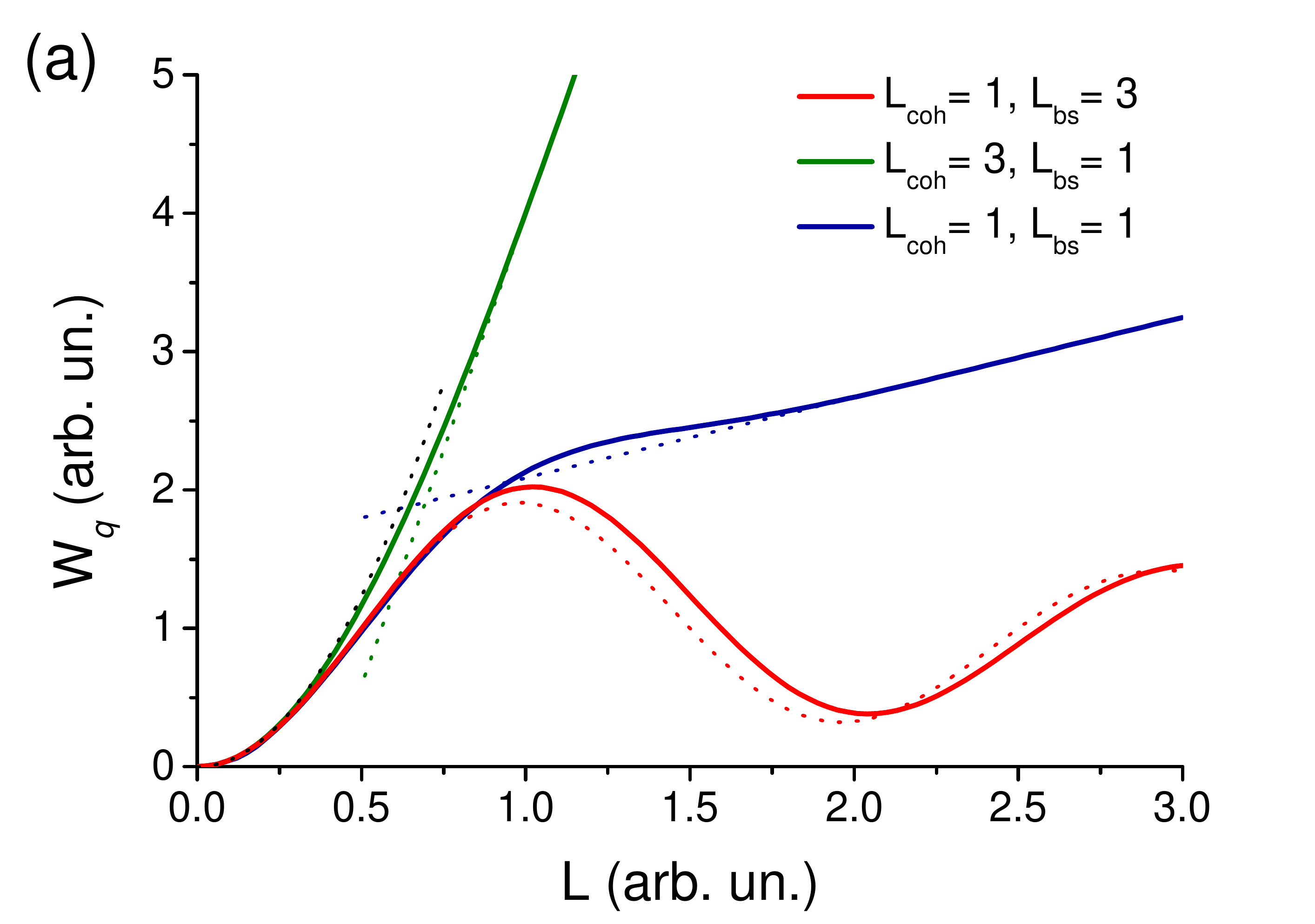}
\includegraphics[width=0.48\linewidth]{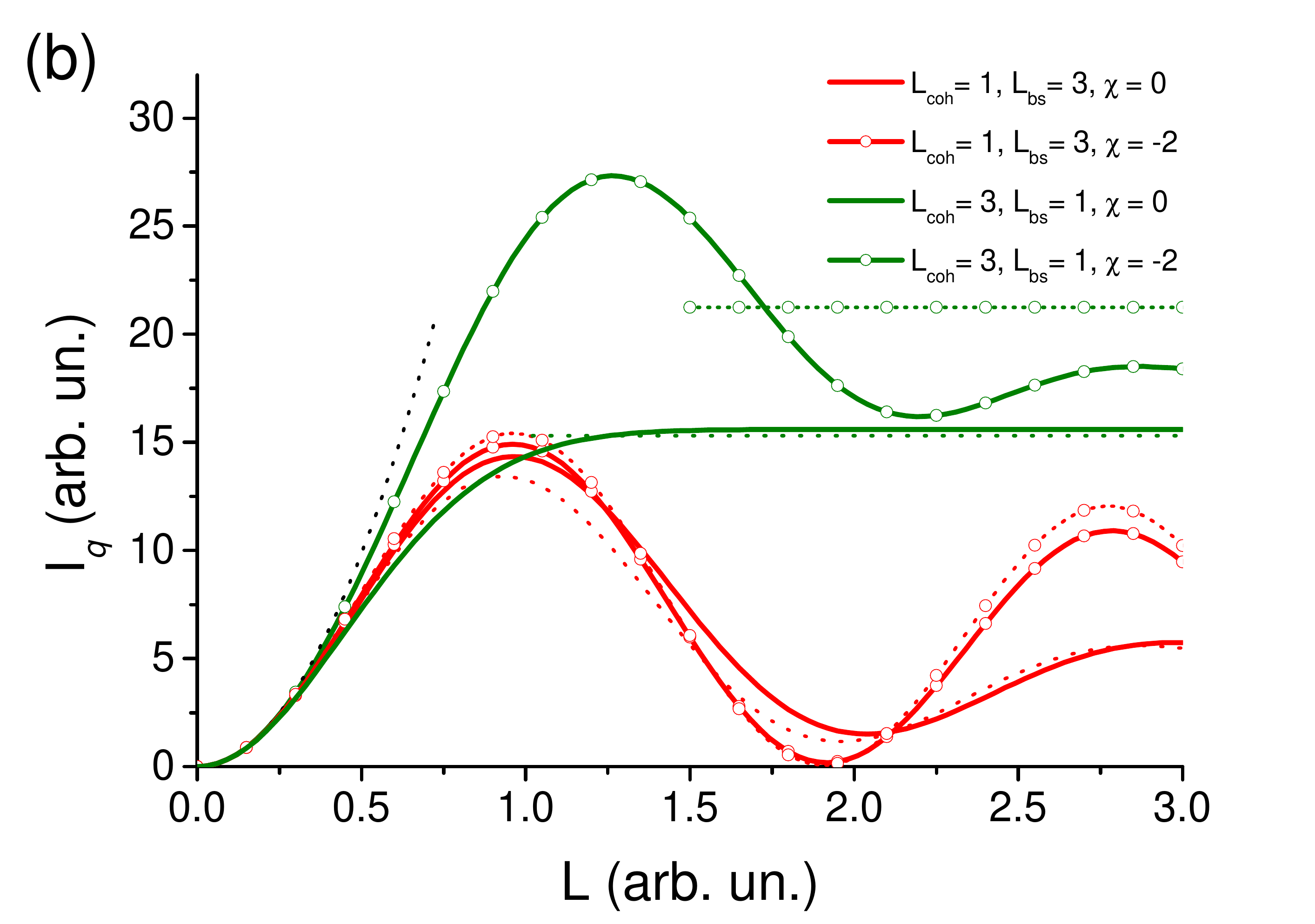}
\caption{Harmonic (a) energy and (b) intensity as functions of the propagation distance. Different colours correspond to different regimes of generation: the short propagation regime (black), the phase-matching defined regime (red) for $L_\mathrm{coh}=1$ and $L_\mathrm{bs}=3$, the blue-shift defined regime for $L_\mathrm{coh}=3$ and $L_\mathrm{bs}=1$ (green), for $L_\mathrm{coh}=1$ and $L_\mathrm{bs}=1$ (blue). For the harmonic energy~(a) solid lines present the exact results~\eqref{tot_energy}, and the dashed lines show the approximate results~\eqref{main_tot_energy_short_L1}, \eqref{main_tot_energy_long_L2} and \eqref{main_tot_energy_long_L5}. 
For the harmonic intensity~(b) solid lines present the exact results~\eqref{int_theory}, and the dashed lines show the approximate results~\eqref{int_short_L}, \eqref{int_all_L} and \eqref{int_long_L}; lines with symbols present the results for the chirped harmonic pulse with $\chi=-2$. The values of the microscopic response $\mathcal{F}_0$ and the pulse duration $\tau$ are the same for all curves, as well as the arbitrary units measuring all lengths.}
\label{Fig_appendix}
\end{figure}

The behaviour of the harmonic total energy $W_q$ under different conditions is illustrated in Fig.~\ref{Fig_appendix}(a). The exact results~\eqref{tot_energy} for several values of the coherence length $L_\mathrm{coh}$ and of the blue-shift length $L_\mathrm{bs}$ are shown with solid curves. The approximate solutions are shown with dashed lines. The generation at short propagation distances demonstrates a quadratic energy growth in agreement with~\eqref{main_tot_energy_short_L1} (black). For $L_\mathrm{coh}=1$, $L_\mathrm{bs}=3$ (red) one can see oscillations characteristic to the phase-matching defined regime~\eqref{main_tot_energy_long_L2}. In the blue-shift defined regime we consider two situations: $L_\mathrm{coh}=3$ and  $L_\mathrm{bs}=1$ (green), where the energy grows fast, according to Eqs.~\eqref{main_tot_energy_long_L5} and \eqref{main_D1}; and $L_\mathrm{coh}=1$ and $L_\mathrm{bs}=1$ (blue), where the harmonic energy grows slower in accordance with Eqs.~\eqref{main_tot_energy_long_L5} and \eqref{main_D}. The important conclusion from Eq.~\eqref{main_tot_energy_long_L5} is that the rate of the linear growth is maximal when $L_\mathrm{coh}$ is infinity, i.e. the free electrons compensate the medium dispersion in the centre of the pulse. Note that the value of such compensation was highlighted in Ref.~\cite{Rundquist1998}.

\subsection{XUV intensity \textit{vs} propagation distance}
Here we study the XUV intensity at the exact harmonic frequency $\omega=q \omega_0$ as a function of the propagation distance. 

The harmonic phase dependence on the laser intensity leads to a chirp of the harmonic pulse. Assuming the linear dependence of the harmonic phase on the laser intensity $\phi_q(t)=-\alpha_q \tilde{I}_\mathrm{las}(t)$~\cite{HHG_phase1, Balcou1997, Gaarde_HHG_HFM}, along with the quadratic temporal modulation of the laser intensity near the pulse maximum
$$
\tilde{I}_\mathrm{las}(t) \approx {I}_\mathrm{las}- {I}_\mathrm{las} \frac{\kappa_0^2 t^2}{ 2 \tau^2} \, ,
$$
where $I_\mathrm{las}$ is the peak laser intensity and $\kappa_0$ is given by Eq.~\eqref{kappa0}, we find that the chirp near the peak of the pulse is linear and that the lineshape of the microscopic response can be described by Eq.~\eqref{Gauss}, where $\Delta \omega$ is given by Eq.~\eqref{kappa} with     
\begin{equation}
\label{eq:kappa_G}
\kappa= \kappa_0 \sqrt{ 1-i 2 \alpha_q I_\mathrm{las} }=\kappa_0 \sqrt{1+i \chi} \, ,
\end{equation}
with $\chi=-2 \alpha_q I_\mathrm{las}$. Note that negative $\chi$ corresponds to the negative chirp (the frequency decreases in time), which is the case for HHG. 

Similar to Eq.~\eqref{main_tot_energy_short_L1}, applying Taylor expansion over $L$ in~\eqref{int_theory}, we explore different regimes for the behaviour of the harmonic intensity with propagation and summarise it in Table~\ref{tab:harm_int_limits}. \\
\begin{table}[h!]
\newlength{\equationlengthtabletwo}
\setlength{\equationlengthtabletwo}{0.7cm}
\renewcommand\thetable{2}
\centering
\begin{tabular}{ | c | c | c  c| }
\hline
Regime & Condition & Harmonic intensity & \\ 
\hline
$\begin{array}{c}
\\
\text{Short} \\ \text{propagation} \\ \text{distances}
\end{array}$
&
\begin{tabular}{r}
\\ for
$
L\ll \min \left\{L_\mathrm{coh}, L_\mathrm{bs} \sqrt{1+\chi^2} \right\}
$
\\
\begin{minipage}{1.5cm}
\begin{equation} \label{condition3} \end{equation}
\end{minipage}
\end{tabular}
&
$
\displaystyle
I_q \approx 4 \pi^2 (2 \pi \mathcal{F}_0 )^2 \left( \frac{L}{\lambda_q}\right)^2 
$
&
\begin{minipage}{\equationlengthtabletwo}
\begin{equation} \label{int_short_L} \end{equation}
\end{minipage}
\\
$\begin{array}{c}
\\
\text{Phase-} \\ \text{matching} \\ \text{defined}
\\ \text{ }
\end{array}$
&
for all $L$ under $L_\mathrm{coh} \ll L_\mathrm{bs}$ 
& 
$
\displaystyle
I_q \approx 4 (2 \pi \mathcal{F}_0 )^2 \left( \frac{L_\mathrm{coh}}{\lambda_q}\right)^2
\left\{B^2+1-2 B \cos\left(\frac{\pi L}{L_\mathrm{coh}}-\frac{\pi^2 L^2 \chi}{8\ln(2) L_\mathrm{bs}^2 (1+ \chi^2)}\right) \right\}
$
&
\begin{minipage}{\equationlengthtabletwo}
\begin{equation} \label{int_all_L} \end{equation}
\end{minipage}
\\ 
$\begin{array}{c}
\text{Blue-shift} \\ \text{defined}
\end{array}$
&
$\begin{array}{c}
\text{for }L>L_\mathrm{bs} \sqrt{1+\chi^2} \\
\text{under } L_\mathrm{bs} \left(1+\chi^2 \right)^{1/4} \lesssim L_\mathrm{coh} \\
\end{array}$
&
$
\displaystyle
I_q \approx  \pi (2 \pi \mathcal{F}_0 )^2  8 \ln(2) (L_\mathrm{bs}/ \lambda_q)^2  \exp{-b^2} \sqrt{1+ \chi^2} \left| 1- i b \sqrt{\frac{2 (1+i\chi)}{\pi}}\right|^2 
$
&
\begin{minipage}{\equationlengthtabletwo}
\begin{equation} \label{int_long_L} \end{equation}
\end{minipage}
\\
&&& \\
\hline
\end{tabular}
\caption{Harmonic intensity in different generation regimes.}
\label{tab:harm_int_limits}
\end{table}

\noindent
Here $B=B(L)$ is defined as
$$
B(L)=\exp\left( - \frac{\pi^2 L^2}{8\ln(2) L_\mathrm{bs}^2 (1+\chi^2)}\right) \, . 
$$

In Table~\ref{tab:harm_int_limits} the condition~\eqref{condition3} for the short distance propagation regime depends on the harmonic chirp [defined by~$\chi$~\eqref{eq:kappa_G}], in contrast to condition~\eqref{condition0}. 

For harmonics from the lowest part of the plateau the condition $\alpha_q I_\mathrm{las} \ll 1$ takes place, thus the condition~\eqref{condition3} can be rewritten as
\begin{equation}
L\ll \min \left\{L_\mathrm{coh}, L_\mathrm{bs} \right\} \, ,
\label{condition4}
\end{equation}
and for harmonics near the cut-off we have $\alpha_q I_\mathrm{las} \gg 1$, thus the condition~\eqref{condition3} becomes
\begin{equation}
L\ll \min \left\{L_\mathrm{coh}, 2 \alpha_q I_\mathrm{las}  L_\mathrm{bs} \right\}.
\label{condition5}
\end{equation}

Figure~\ref{Fig_appendix}(b) illustrates the behaviour of the XUV intensity for a given frequency as a function of the propagation length, as well as some approximations found above. One can see that when this behaviour is defined by the phase matching both the total energy [see Fig.~\ref{Fig_appendix}(a)] and the intensity oscillate, and when it is defined by the blue shift, the energy linearly increases with the propagation distance and the intensity saturates. Note that the case $L_\mathrm{coh}=L_\mathrm{bs}$ [blue curve in the graph~(a)] corresponds rather to the blue-shift defined generation, because the energy grows with the distance.

\section{Numerical method}
\label{numerical}
We simulate the generation and propagation of the fields via numerically solving 1D propagation equations for the generating and generated fields with the (nonlinear) polarisation of the medium extracted from the numerical solution of the 3D time-dependent Schr\"odinger equation~(TDSE) in the single-active-electron~(SAE) approximation for a model argon atom at every propagation step. In the TDSE solution the contributions of the longer electronic trajectories are suppressed, so the contribution of the shortest one dominates. See details of the method in~\cite{KhokhlovaStrelkov, Birulia2022}.

As we mentioned above, the coherence length is ill-defined due to the temporal variation of the free-electron contribution to the medium dispersion. We define $L_\mathrm{coh}$ via~\eqref{L_coh_general} using the mismatch $|\Delta k|$ for the central spectral component of the laser. In more details, we first check in the numerical propagation that the harmonic refraction is negligible. Then we find the phase advance $\Delta \varphi$ of the central laser frequency accumulated at the propagation distance $\Delta x$ and calculate $|\Delta k_q|=q |\Delta \varphi|/ \Delta x$ (note, that $|\Delta \varphi|$ increases almost linearly with the propagation length within the first dozens of propagation steps). The mismatch $|\Delta k|$ is defined, in particular, by dispersion of the neutral gas. Its dispersion is reproduced in our calculations fairly well: for real argon under $\lambda_0=\SI{0.8}{\micro m}$ laser wavelength, the ionisation 3.2\% (in the centre of the laser pulse) compensates the neutral gas dispersion; in our simulation this value is 4\%.  

Finally, to calculate $L_\mathrm{bs}$~\eqref{L_bs}, we find the ionisation probability after the pulse $w$ (it almost does not change with the propagation distance for the distances considered here). Then we obtain
\beq{nn}
|n_\mathrm{f}-n_\mathrm{i}| = 4.5 \times 10^{-22} \: w \:  N [\mathrm{cm}^{-3}] \lambda_0^2 [\mu \mathrm{m}] \, ,
\eeq
where $N$ is the initial atomic density and $\lambda_0$ is the driver wavelength. 

The medium density we use throughout is $\SI{3e18}{cm^{-3}}$. Note that the propagated results depend on the product of the length and the density, so the results can be attributed to other densities $N$ via multiplying the propagation distance presented in the results below by $3 \times 10^{18} / N [\mathrm{cm}^{-3}]$.

Below we also use the normalised length scales defined as 
\beq{L_norm}
\tilde L_\mathrm{coh, bs} = L_\mathrm{coh, bs}^{(q)}/\lambda_q \, .
\eeq
This normalisation is convenient because $\tilde L_\mathrm{coh, bs}$ do not depend on the harmonic order $q$, see~(\ref{L_coh}),~(\ref{L_bs}). The tilde symbol appearing with other length scales means similar normalisation.

\section{Results}\label{Results}
\begin{figure}
\centering
\includegraphics[width=0.55\linewidth]{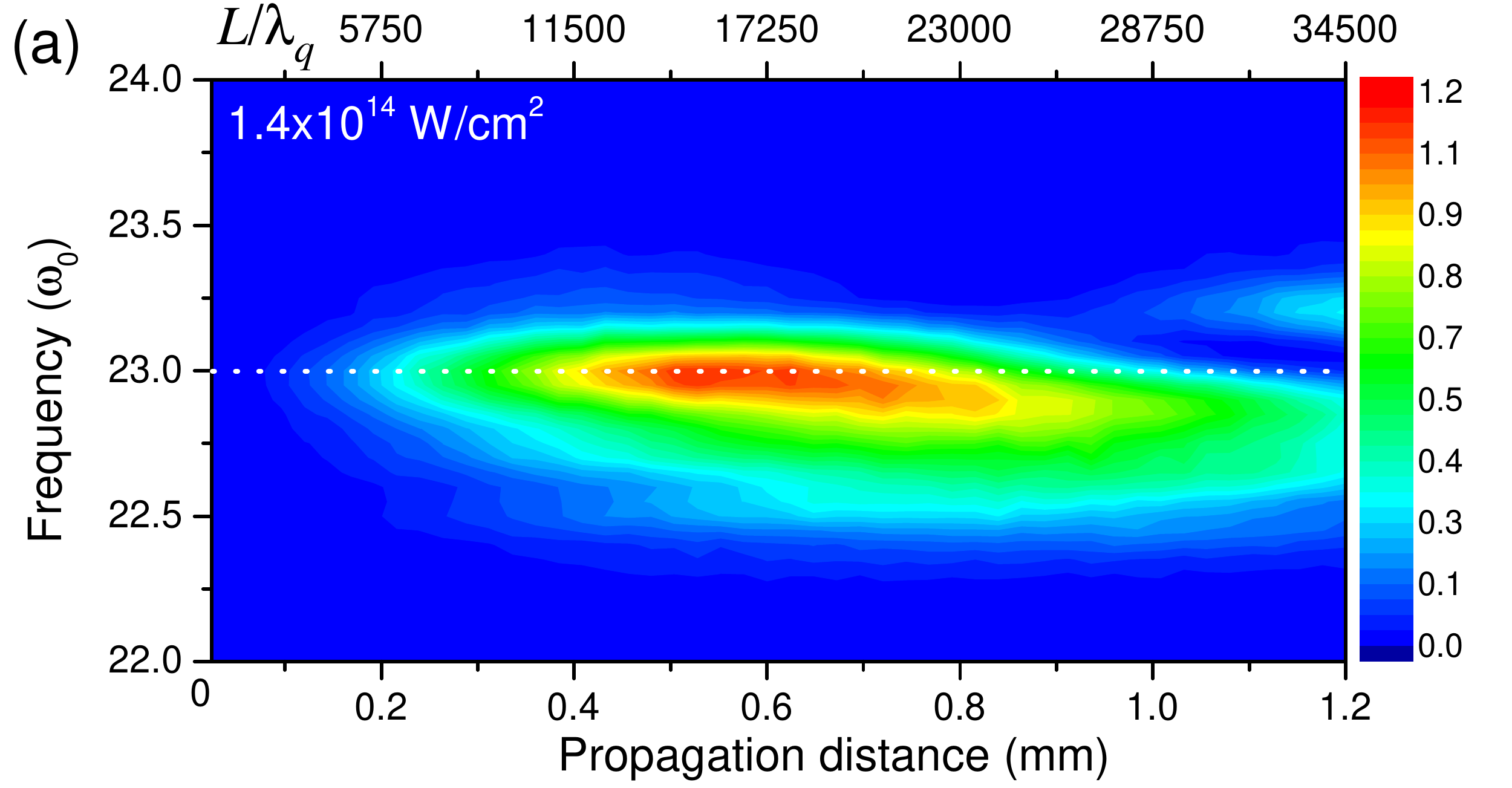}
\vspace{2mm}
\includegraphics[width=0.55\linewidth]{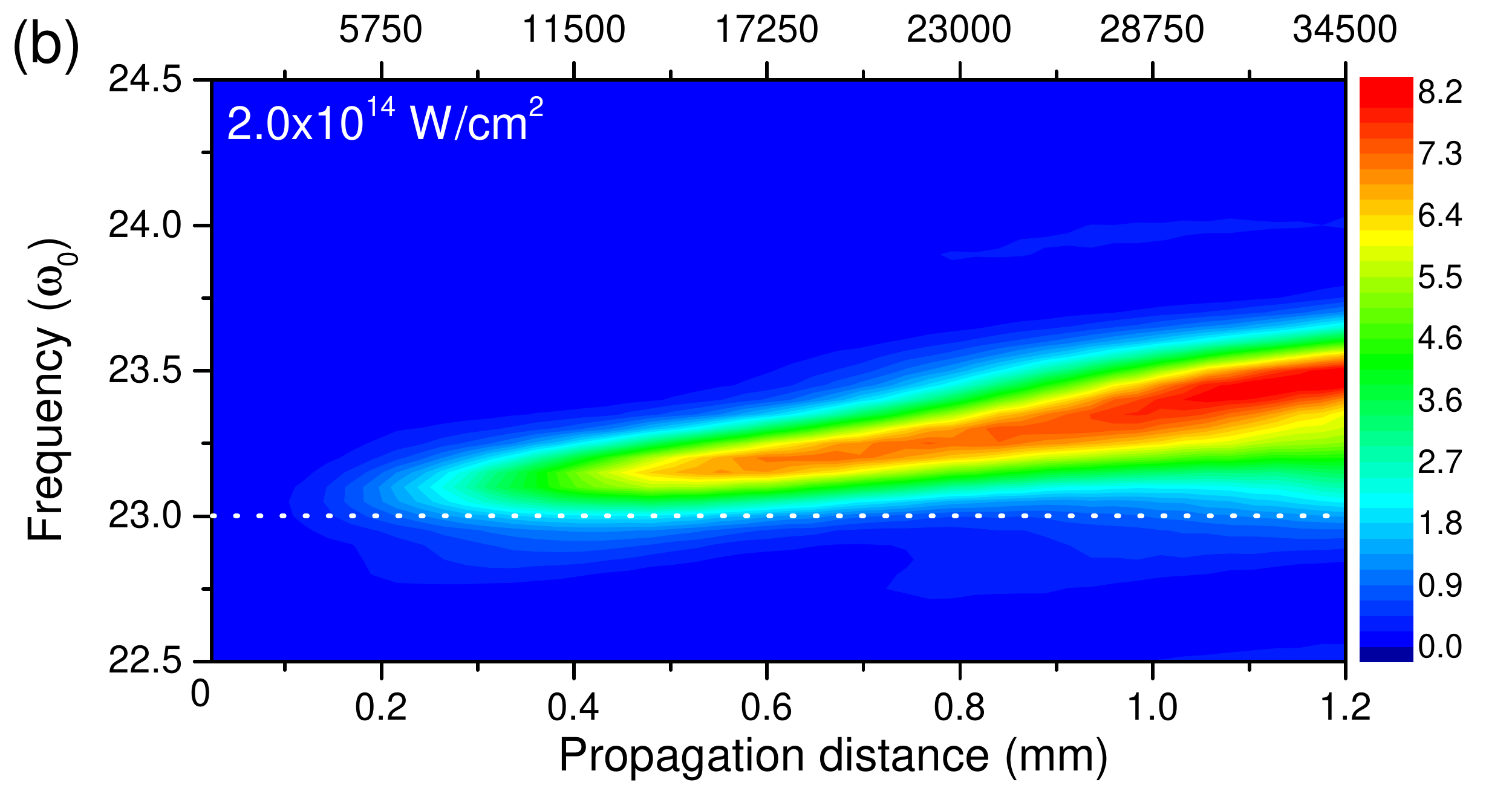}
\vspace{2mm}
\includegraphics[width=0.55\linewidth]{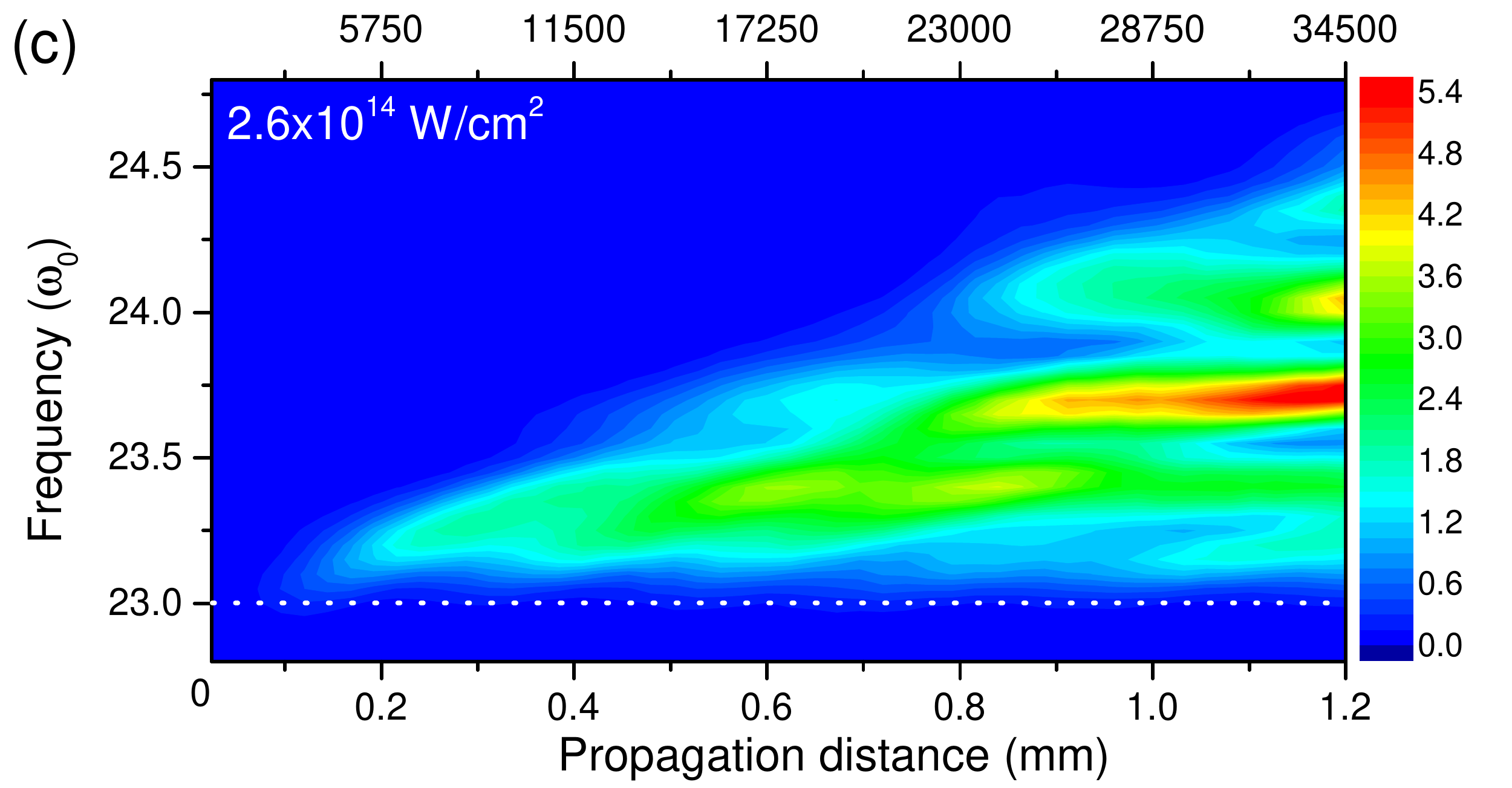}
\vspace{2mm}
\includegraphics[width=0.55\linewidth]{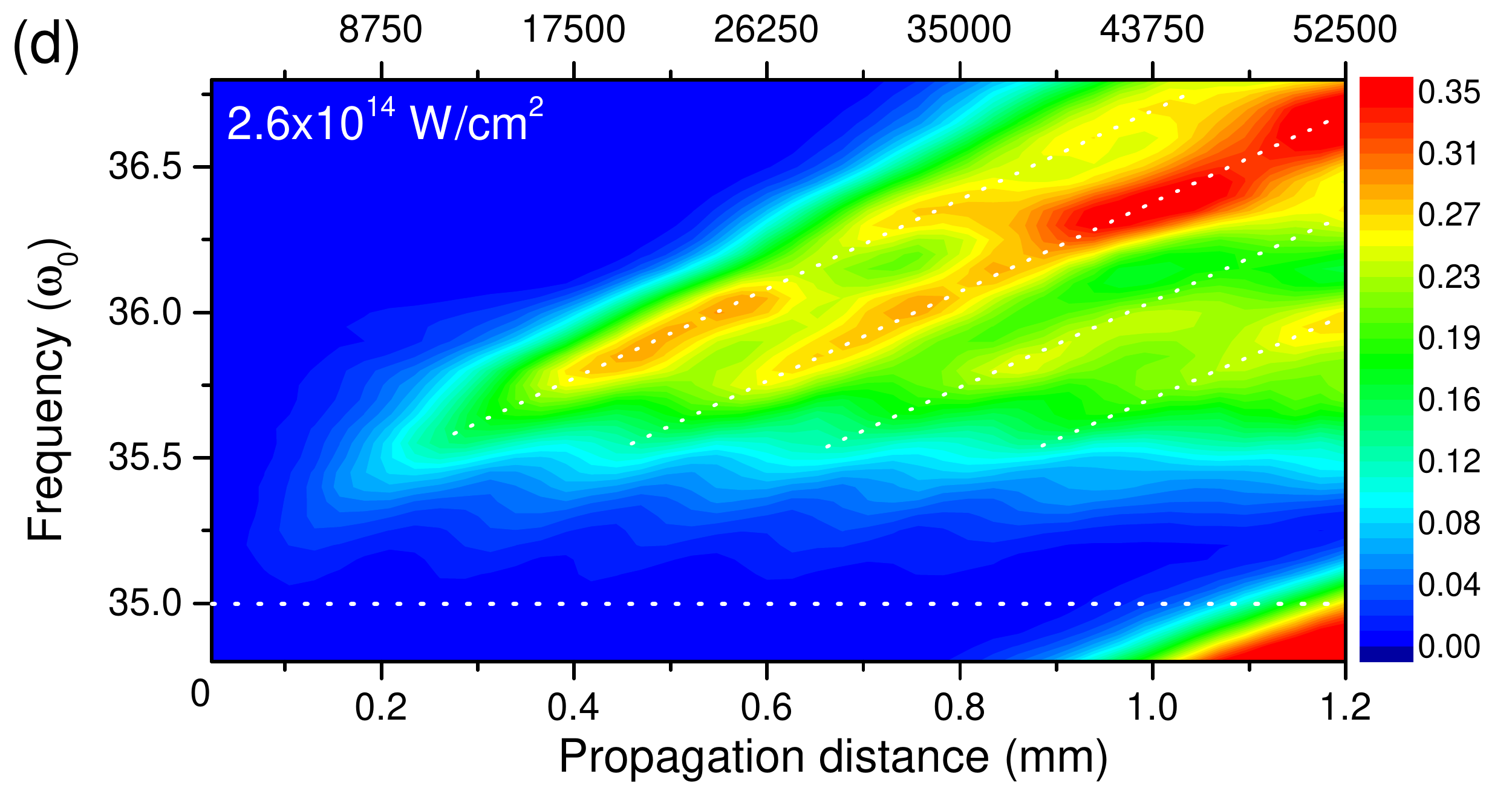}
\caption{Spectra of (a-c) H23 and (d) H35 as functions of the propagation distance. The driving intensities are (a)~$\SI{1.4e14}{W/cm^2}$, (b)~$\SI{2e14}{W/cm^2}$ and (c,d)~$\SI{2.6e14}{W/cm^2}$. The laser pulse duration is \SI{20}{fs}, the laser wavelength is \SI{800}{nm}. Top horizontal axes show the propagation distance in units of harmonic wavelengths.}
\label{spectra}
\end{figure}

Figure~\ref{spectra} presents the simulated spectra in the vicinity of the 23$^\mathrm{d}$ harmonic (H23), see panels (a-c), and H35, see panel~(d), as functions of the propagation distance. One can see two different behaviours of the harmonic spectrum with the propagation distance: for the low laser intensity~(a) the XUV intensity initially grows without notable change of the harmonic linewidth and of its central frequency, and then decreases. For higher laser intensities~(b,c) the XUV intensity grows monotonically, the linewidth and the central frequency increase. This increase is more pronounced for the higher driving intensity~(c). 

In Fig.~\ref{spectra}(b,c) one can see that after propagation over some distance the harmonic line has a typical `foot' shape: a~pronounced blue-shifted peak and a wide pedestal at the red side of the peak. Such harmonic lineshape was observed experimentally (see, for instance, Ref.~\cite{Ellipticity2005}). It can be explained as follows: at the front of the pulse the XUV frequency is blue-shifted due to the dependence of the harmonic phase on the laser intensity. At the peak of the pulse it is blue-shifted due to plasma-induced blue shift of the driver. At the falling edge of the pulse this blue shift is compensated by the XUV red shift, which appears due to the harmonic phase dependence on the laser intensity, leading to the wide pedestal at the red side of the peak.   

\begin{figure}
\includegraphics[width=0.48\linewidth]{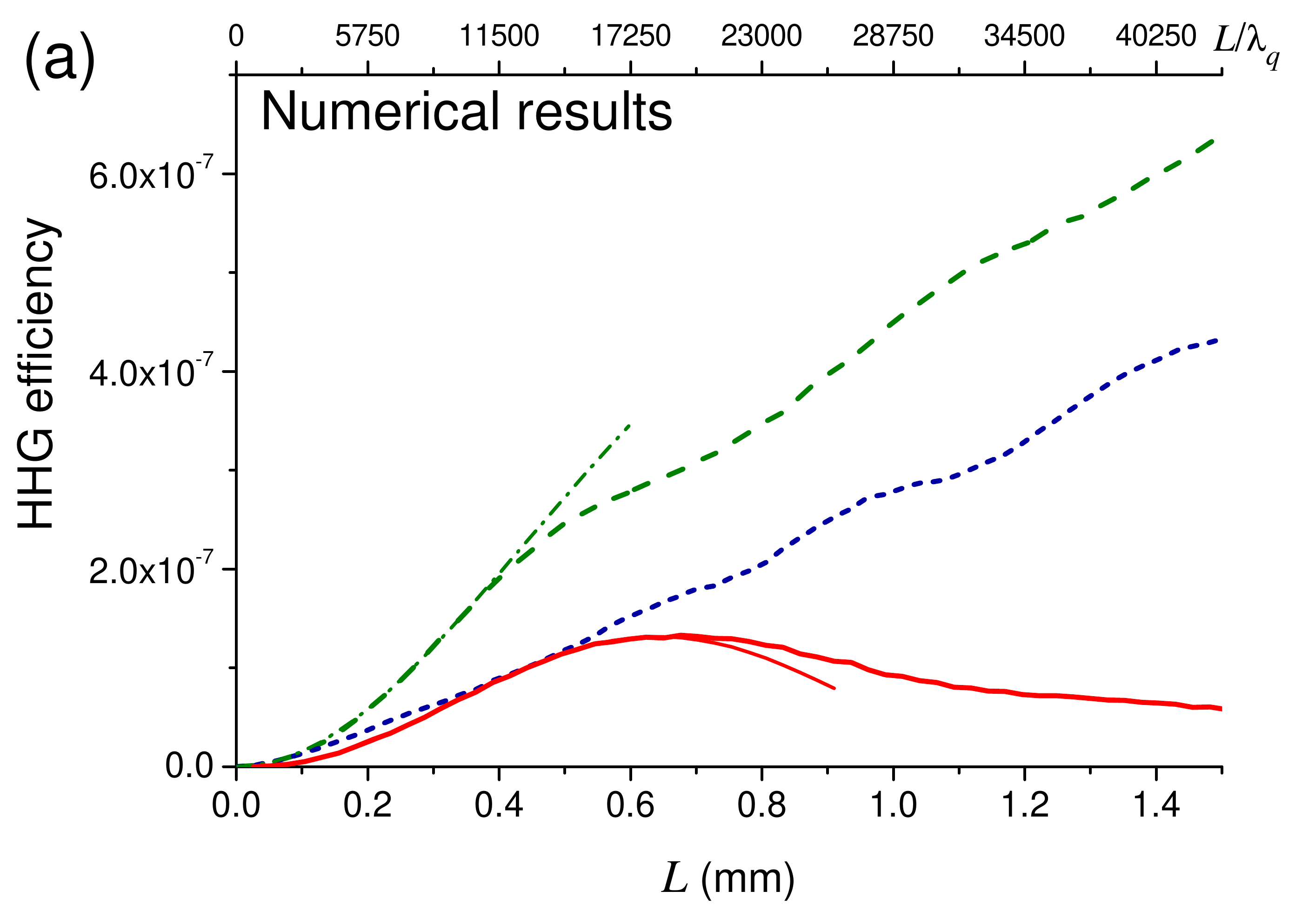}
\includegraphics[width=0.48\linewidth]{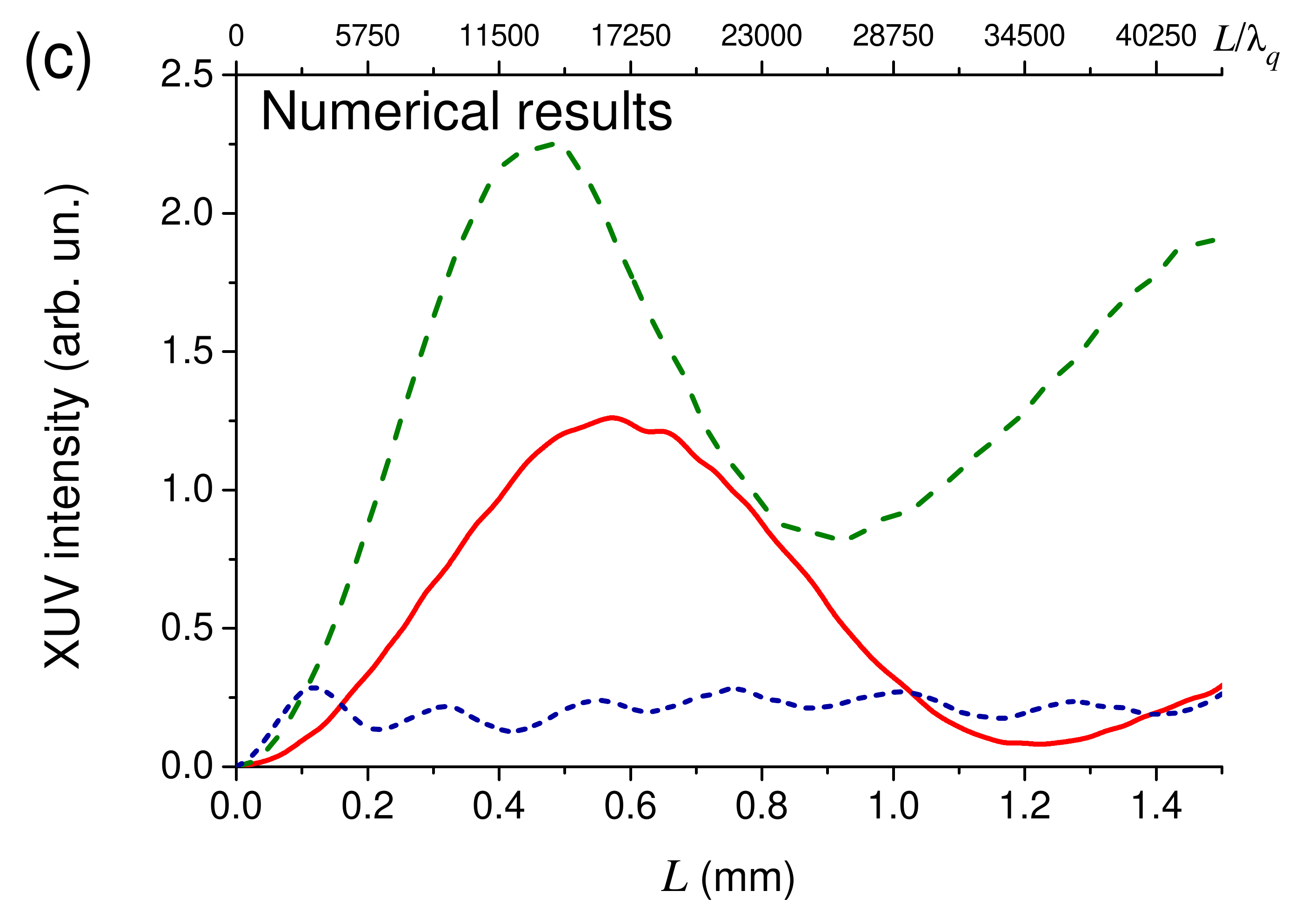}
\includegraphics[width=0.48\linewidth]{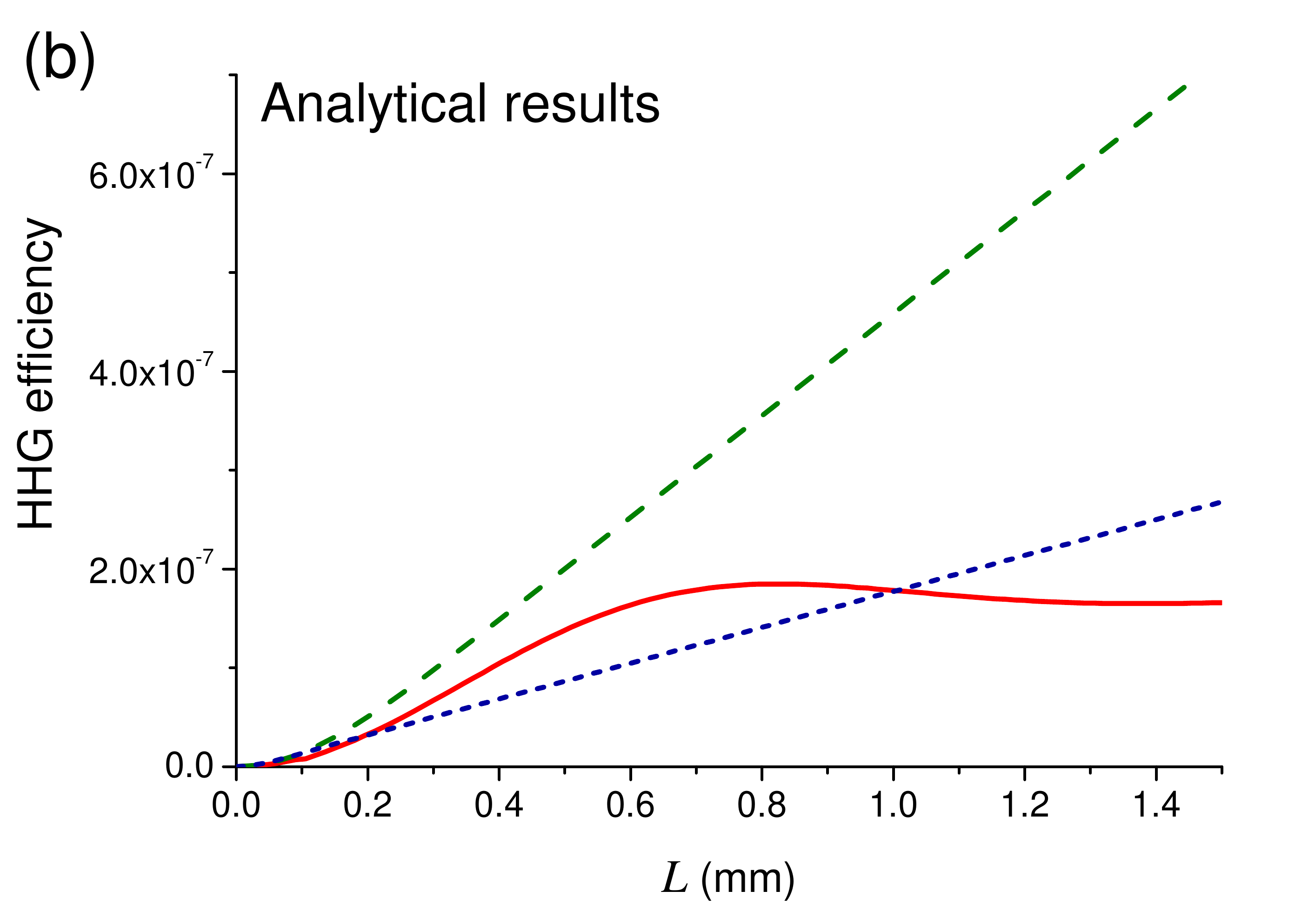}
\includegraphics[width=0.48\linewidth]{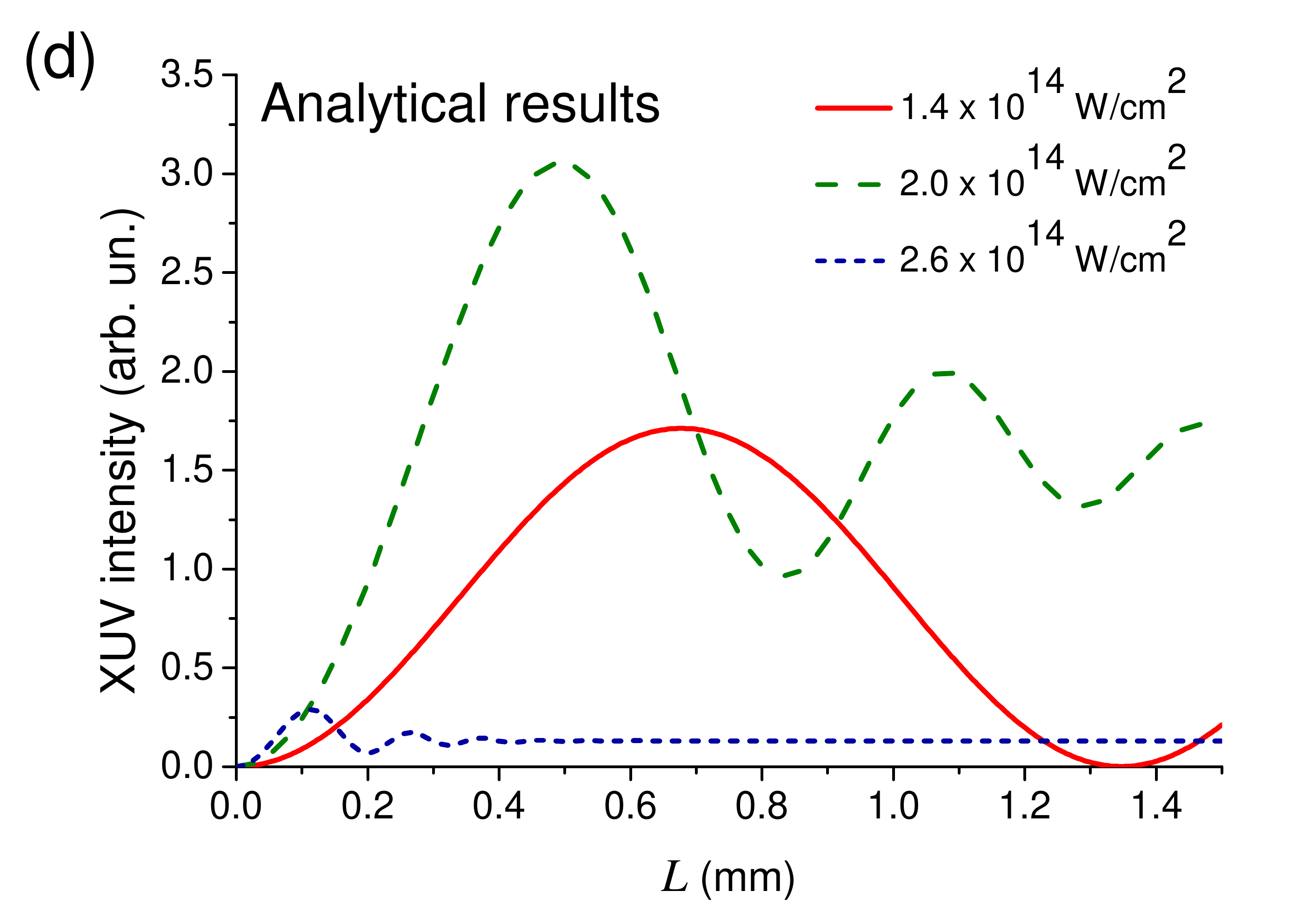}
\caption{(a,b)~H23 generation efficiency and (c,d)~spectral intensity at $\omega=23 \omega_0$ as functions of the propagation distance for different driving intensities, calculated (a,c)~numerically and (b,d)~analytically. Different colours show the results for the different laser intensities marked in panel~(d), which correspond to the same colour-coded generation regimes as in Fig.~\ref{Fig_appendix}. Other parameters are the same as in Fig.~\ref{spectra}. The thin lines in panel~(a) present the HHG efficiency approximations~\eqref{cos_approximation} with red and~\eqref{Pade_approximation} with green.}
\label{H23_theory_vs_num}
\end{figure}

Figure~\ref{H23_theory_vs_num}(a,b) presents the harmonic efficiency for H23 calculated as the total energy of the harmonic (XUV spectral intensity integrated from $\omega=22.5 \omega_0$ to $\omega=24.5 \omega_0$) divided by the laser pulse energy. The efficiency also starts growing quadratically, and later either saturates or decreases for the weakest laser intensity, or grows linearly for the higher ones. 
 
Figure~\ref{H23_theory_vs_num} also presents the comparison of the numerical~(a,c) and theoretical~(b,d) results for both the H23 generation efficiency and the XUV intensity. Note that the slice of the results shown in Fig.~\ref{spectra}(a-c) for the XUV frequency $\omega=23 \omega_0$ is presented in Fig.~\ref{H23_theory_vs_num}(c). Here one can see that the XUV intensity initially grows quadratically with the propagation distance, then switches to linear growth, and later either decreases to almost zero (for the weakest laser intensity) or saturates (for higher ones) continuing to oscillate. For the analytical study we use $\tilde L_\mathrm{coh}$ and $\tilde L_\mathrm{bs}$ calculated using the values of $\Delta k$, $n_\mathrm{f}$ and $n_\mathrm{i}$ extracted from our numerical calculations: $\tilde L_\mathrm{coh}=19750$ and $\tilde L_\mathrm{bs}=24500$ for the laser intensity $\SI{1.4e14}{W/cm^2}$, $\tilde L_\mathrm{coh}=34250$ and $\tilde L_\mathrm{bs}=5870$ for $\SI{2.0e14}{W/cm^2}$, and $\tilde L_\mathrm{coh}=5270$ and $\tilde L_\mathrm{bs}=2050$ for $\SI{2.6e14}{W/cm^2}$; for every laser intensity we use the harmonic microscopic responses $\mathcal{F}_0$ found numerically for this intensity; for values of $\alpha$ we use numerical results of Ref.~\cite{Khokhlova}: $\alpha_{23}$ equals to 13, 2, and 1 for intensities $1.4 \times 10^{14}$, $2.0 \times 10^{14}$, $2.6 \times 10^{14}$ W/cm$^2$, correspondingly. Fig.~\ref{H23_theory_vs_num} shows that the analytical theory reproduces numerical results reasonably well. 

Now we focus on the short propagation distances. Namely, we find the distance $L_\mathrm{num}$ where the numerical HHG efficiency changes quadratic growth to the linear one, and compare it with theoretical values of $L_\mathrm{coh}$ and $L_\mathrm{bs}$. To do this, we first fit the numerical signal to reconstruct $L_\mathrm{num}$ from this fit. Unfortunately, the general fit (the one which adequately describes the signal in both phase-matching and blue-shift defined regimes) turns out to be very unstable. To overcome this problem, we use different fits in different regimes. When the calculated signal demonstrates a pronounced decrease after initial increase (this means that the HHG is phase-matching defined) the signal can be fitted (see Eq.~\eqref{main_tot_energy_long_L2}) as
\begin{equation}
W_q \propto 
\left\{ 1 - \cos\left(\frac{\pi L}{2 L_\mathrm{num}}\right) \right\} \, ,
    \label{cos_approximation}
\end{equation}
where $L_\mathrm{num}$ should be close to $L_\mathrm{coh}/2$. If the signal keeps growing, it is more convenient to use the fitting based on the Pad\'e approximation, see Eq.~\eqref{Pade_approximation1} in Appendix:
\begin{equation}
W_q \propto 
\frac{L^2}{1+L^2/(6 L_\mathrm{num}^2)} \, ,
\label{Pade_approximation}
\end{equation}
where
$L_\mathrm{num}$ should be close to 
$L_\mathrm{bs} L_\mathrm{coh} \sqrt{2}/\pi/\sqrt{L_\mathrm{bs}^2+\left(L_\mathrm{coh} \sqrt{2}/\pi \right)^2} $. In particular, in the blue-shift defined regime ($L_\mathrm{bs} \ll L_\mathrm{coh}$) from the latter equation we have $L_\mathrm{num}\approx{L_\mathrm{bs}}$. The two approximations are shown in Fig.~\ref{H23_theory_vs_num}(a).

\begin{figure}
\centering
\includegraphics[width=0.6\linewidth]{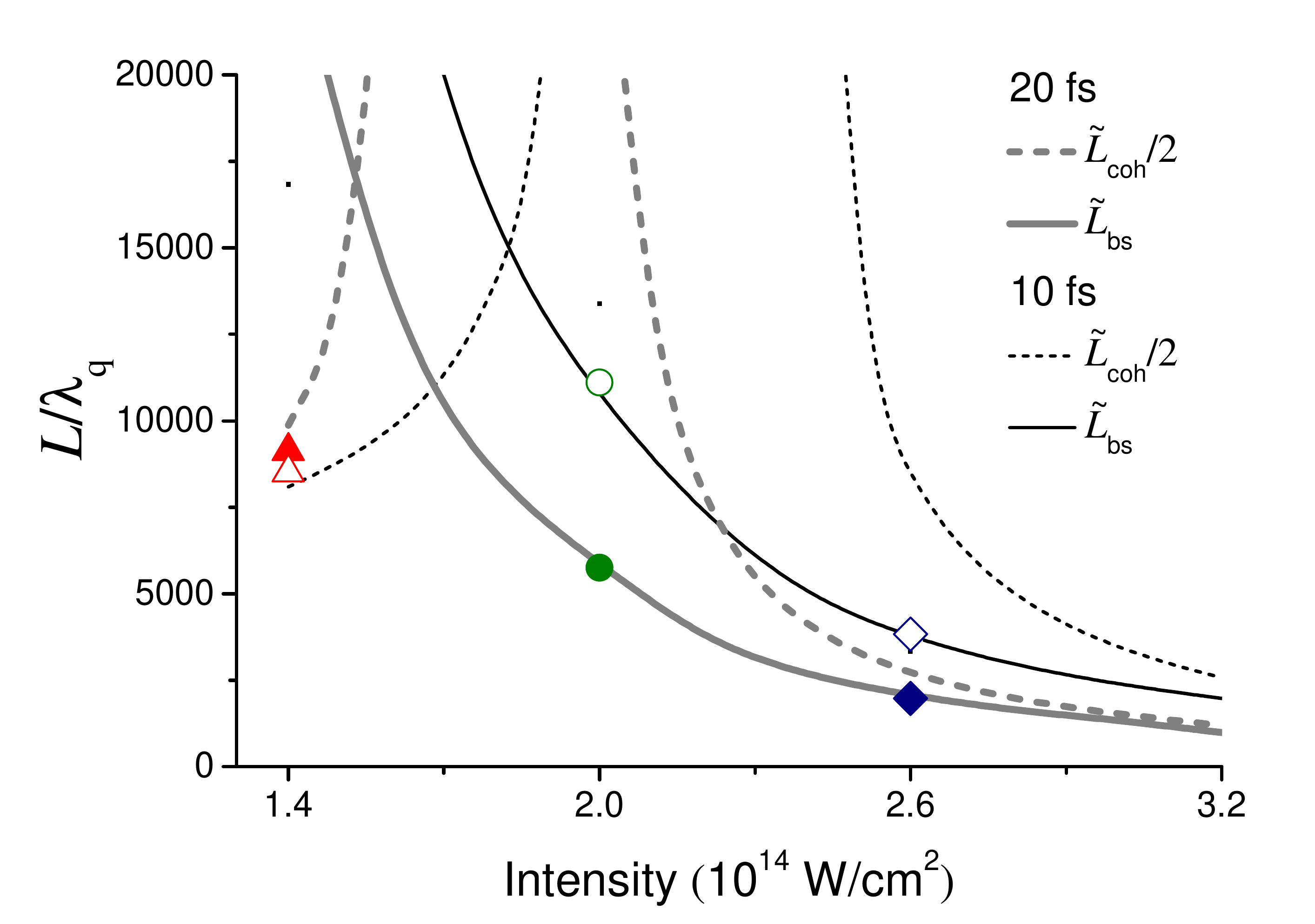}
\caption{The normalised length scales~(\ref{L_norm}) describing macroscopic response of H23 as functions of the driving intensity: the blue-shift length $\tilde L_\mathrm{bs}$, the coherence length $\tilde L_\mathrm{coh}$, and the length of quadratic intensity increase $\tilde L_\mathrm{num}$ extracted from the fits of the numerical propagation simulations (symbols). Results are shown for the pulse durations \SI{20}{fs} and \SI{10}{fs}.}
\label{L_vs_int}
\end{figure}

In Fig.~\ref{L_vs_int} we show $\tilde L_\mathrm{num}^{(23)}$ found for different laser intensities and pulse durations, as well as $\tilde L_\mathrm{coh}$ and $\tilde L_\mathrm{bs}$ calculated via Eq.~\eqref{L_coh} and \eqref{L_bs} using values of $\Delta k$, $n_\mathrm{f}$ and $n_\mathrm{i}$ retrieved from our numerical calculations, see Sec.~\ref{numerical}. 
We see that $\tilde L_\mathrm{coh}$ first grows with the laser intensity and then decreases. For certain laser intensity (specific for a given pulse duration, for a shorter pulse it is higher) $ \tilde L_\mathrm{coh}$ is infinity because the neutral atom dispersion is compensated by the plasma dispersion in the centre of the pulse. $\tilde L_\mathrm{bs}$ decreases with the laser intensity because the ionisation degree grows. For high laser intensities $\tilde L_\mathrm{bs} \approx \tilde L_\mathrm{coh}/2$ in agreement with Eqs.~\eqref{L_coh} and~\eqref{L_bs}. (In more details, for high intensity the ionisation is relatively deep, so that $|n_f-n_i| \gg |n_i-1|$, thus $\Delta n \approx |n_f-n_i|/2$.) Comparing $\tilde L_\mathrm{coh}/2$ and $\tilde L_\mathrm{bs}$ with $\tilde L_\mathrm{num}^{(23)}$, one can notice that the latter is close to the shortest of $\tilde L_\mathrm{coh}/2$ and $ \tilde L_\mathrm{bs}$.

\begin{figure}
\centering\includegraphics[width=0.6\linewidth]{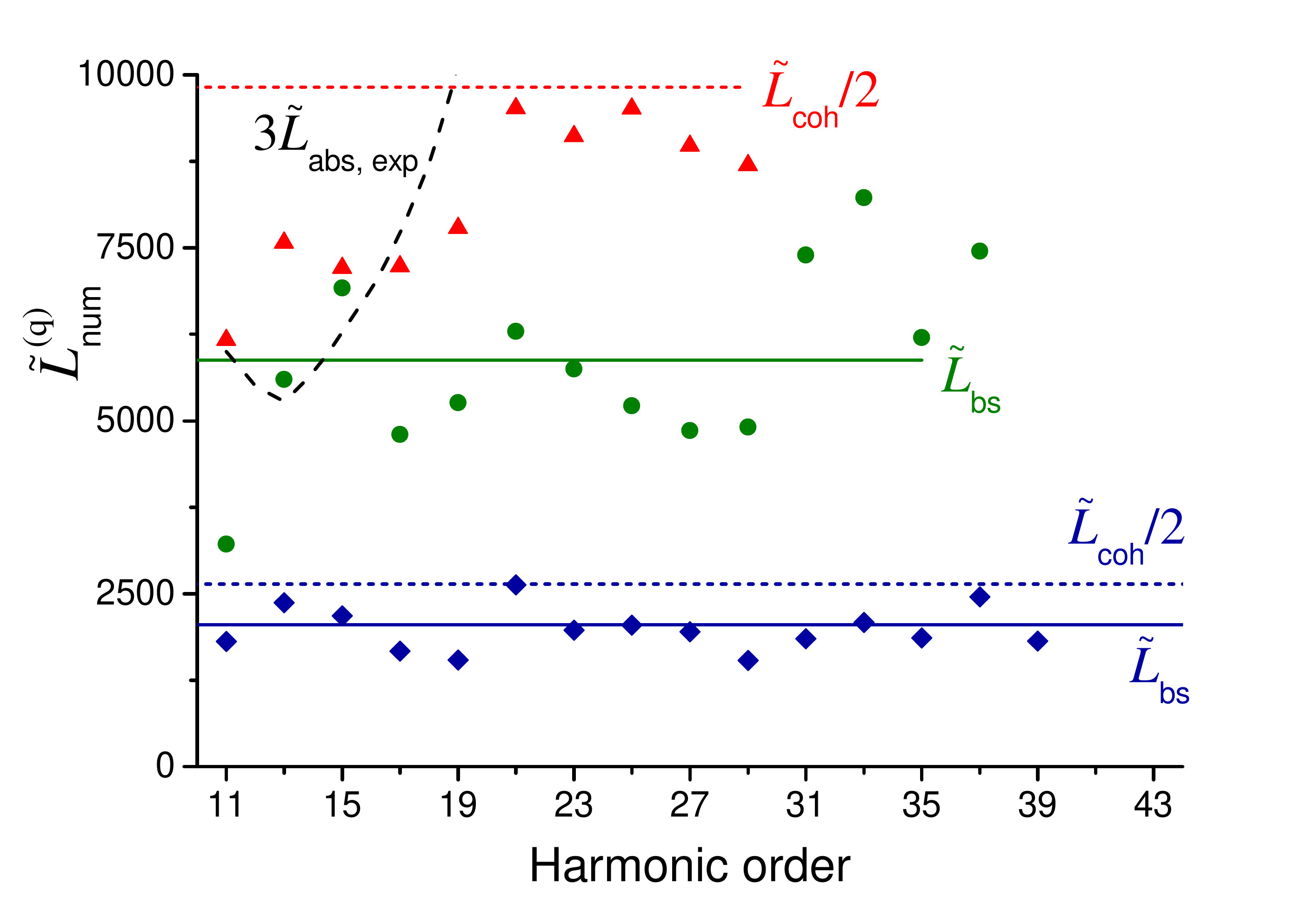}
\caption{$\tilde L_\mathrm{num}^{(q)}$ as a function of the harmonic order $q$ for different driving intensities and driving pulse duration \SI{20}{fs}. Length scales limiting the harmonic quadratic growth are shown by dashed ($L_\mathrm{coh}/2$) and solid ($L_\mathrm{bs}$) lines. Black dashed curve presents the tripled normalised absorption length found for the density $N=3 \times 10^{18}$ cm$^{-3}$ of argon from experimental data~\cite{CXRO}.}
\label{L_num_vs_q}
\end{figure}

Fig.~\ref{L_num_vs_q} presents $L_\mathrm{num}^{(q)}$ as a function of the harmonic order for the three laser intensities. Moreover, we show $\tilde L_\mathrm{coh}/2$ and $\tilde L_\mathrm{bs}$ for the highest laser intensity (when these lengths are close to each other), and the shortest of those two lengths for other intensities. We see that values of $\tilde L_\mathrm{num}^{(q)}$ are similar for different harmonics (except the lowest ones, see the next paragraph), so the limiting mechanism (phase matching or blue shift) is common for different harmonic orders. 

In Fig.~\ref{L_num_vs_q} we also show experimental data on the tripled absorption length (this scale characterises the harmonic signal saturation due to absorption~\cite{Constant}), which illustrates that for the lowest laser intensity (\SI{1.4e14}{W/cm^2}, red triangles) the generation of harmonics below H21 is defined by the absorption. Note that several points above this limit can be attributed to some underestimation of the absorption in our simulations based on the TDSE solution in the SAE approximation for the atomic response. The border of the absorption-defined spectral region becomes lower with the increasing laser intensity and for the intensity \SI{2.6e14}{W/cm^2} (blue diamonds) this harmonic group vanishes.

Note that comparing Figs.~\ref{spectra} and~\ref{L_num_vs_q}, we see that for both cases, $\tilde L_\mathrm{coh} \gg \tilde L_\mathrm{bs}$ (laser intensity $\SI{1.4e14}{W/cm^2}$) and $\tilde L_\mathrm{coh}/2 \approx \tilde L_\mathrm{bs}$ ($\SI{2.6e14}{W/cm^2}$), H23 demonstrates similar features, which are typical for $L_\mathrm{num}^{(q)}$ defined by $L_\mathrm{bs}$: central frequency of the harmonic varies with propagation (see Fig.~\ref{spectra}) eliminating negative interference of the XUV generated at different propagation distances, and thus leading to linear growth of the generation efficiency, see Fig.~\ref{H23_theory_vs_num}(a,b). So the harmonic generation under high laser intensities should be attributed to the blue-shift defined regime.

\begin{figure*}
\centering\includegraphics[width=0.98\linewidth]{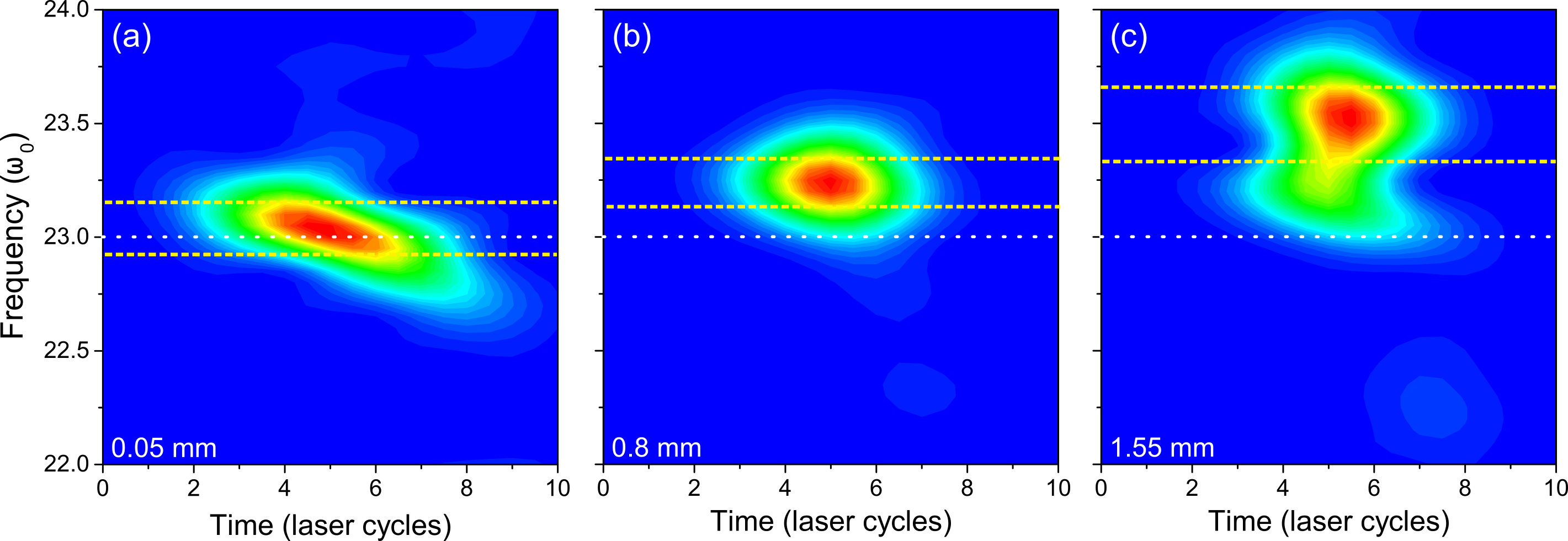}
\caption{Wavelet image of H23 after (a)~\SI{0.05}{mm}, (b)~\SI{0.8}{mm}, and (c)~\SI{1.55}{mm} of propagation. The peak laser intensity is $\SI{2e14}{W/cm^2}$, the pulse duration is \SI{20}{fs}. Colour scales at the graphs are different. White dotted lines show the exact harmonic frequency, while yellow dashed lines mark the same arbitrary intensity level.}
\label{Wavelet}
\end{figure*}
Figure~\ref{Wavelet} describes the temporal dynamics of the H23 spectrum for three different propagation distances. The pulse duration and the intensity are chosen so that the free electrons compensate the neutrals dispersion in the centre of the pulse, so $L_\mathrm{coh}$ is infinite, see Fig.~\ref{L_vs_int}. For the very short distance [panel~(a)] the harmonic field is similar to the atomic response. The harmonic is emitted for a rather long time and it is chirped. For the longer propagation distance [panel~(b)] the field is temporally confined near the centre of the pulse because the phase matching at its edges is poor. Similar temporal confinement was studied in~\cite{Strelkov_2008, Kazamias2011}. The emission at even longer propagation distances [panel~(c)] is shifted to higher frequencies due to the laser blue shift, thus it does not interfere with the lower-frequency field emitted at the shorter propagation distances. As a result, the total field is temporary confined but its bandwidth linearly grows with propagation. This growth continues up to overlapping of neighbour harmonic lines or up to propagation distance close to the absorption length.

\begin{figure}
\centering\includegraphics[width=0.5\linewidth]{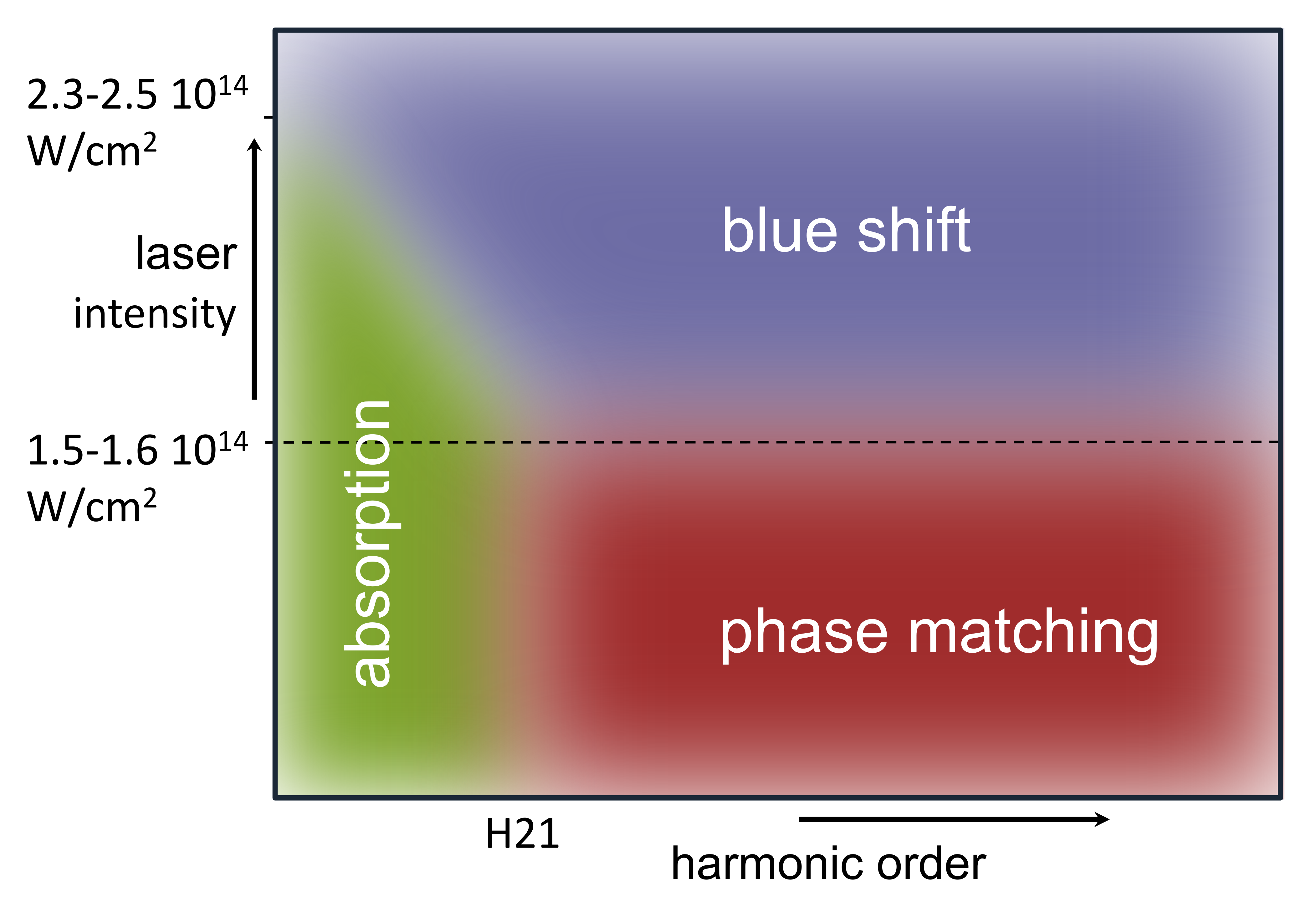}
\caption{Diagram for different mechanisms defining behaviour of HHG signal in argon. Approximate intensities, describing change of the HHG regimes are given for the pulse duration of \SI{20}{fs}.}
\label{Conclusion}
\end{figure}

\section{Conclusions}
Summarising, in this paper we study the macroscopic HHG signal as a function of the propagation distance in argon in 1D geometry. The behaviour of the signal is defined by the shortest of the length scales $L_\mathrm{abs}^{(q)}$, $L_\mathrm{coh}^{(q)}$, $L_\mathrm{bs}^{(q)}$. Correspondingly, we find that the macroscopic behaviour of the HHG signal can be characterised with one of the three regimes schematically presented in Fig.~\ref{Conclusion}. For the lowest harmonics $L_\mathrm{abs}$ is the shortest length scale and their generation is limited by the XUV absorption. In this case the HHG efficiency and XUV intensity at a given frequency first grow quadratically and then saturate after the length of a few $L_\mathrm{abs}$. For the higher harmonics and relatively low laser intensities $L_\mathrm{coh}$ is the shortest length scale. In this regime both HHG efficiency and intensity grow quadratically up to $L_\mathrm{coh}/2$ and decrease after passing $L_\mathrm{coh}$. For higher laser intensities (the changeover intensity is higher for the shorter pulse duration) $L_\mathrm{bs}$ is the shortest length. In this regime the behaviour of the XUV intensity is complicated~--- it grows quadratically up to the shortest of $L_\mathrm{coh}$ and a few $L_\mathrm{bs}$, see Eq.~\eqref{condition3}, and then oscillates or saturates depending on the $L_\mathrm{bs}/L_\mathrm{coh}$ ratio and the harmonic order. The behaviour of the HHG efficiency is simpler~--- first it grows quadratically up $L_\mathrm{bs}$ and then grows linearly (see Eq.~\eqref{main_tot_energy_long_L5}) due to the linear increase of the harmonic bandwidth. The rate of the efficiency growth is the highest for a certain combination of the peak laser intensity and the pulse duration when the free electrons compensate the neutrals dispersion near the centre of the pulse. This growth continues up to the overlap of the neighbouring harmonic spectral lines or up to the limit given by the absorption.

\section*{Appendix}
\subsection{Asymptotic behaviour of the harmonic energy with propagation distance}
Introducing the normalised propagation length from the exponent in~\eqref{tot_energy}
\begin{equation}
l= \frac{\pi L}{4\sqrt{ \ln(2)} L_\mathrm{bs}} \, ,
    \label{l}
\end{equation}
we rewrite Eq.~\eqref{tot_energy} as
\begin{equation}
W_q = \sqrt{\frac{32}{\pi}} \Delta \omega \overline{I^{(q)}} 
\left\{    
\sqrt{\pi}     \Re{ \left[ (l - i b) \erf{(l - i b)}\right] } + \sqrt{\pi} b \ \mathrm{erfi} {(b)} + e^{b^2} \left[ \cos{(2bl)} e^{-l^2} -1  \right]    
\right\} \, ,
   \label{tot_energy1}
\end{equation}
where $b$ is given by Eq.~(\ref{main_b}). This parameter is proportional to the ratio of $L_\mathrm{bs}$ and $L_\mathrm{coh}$. Substituting 
$$
\overline{I^{(q)}}= \pi 8 \ln (2) (2 \pi \mathcal{F}_0)^2 (L_\mathrm{bs}/ \lambda_q)^2 e^{-b^2} \, ,
$$
we have 
\begin{equation}
W_q = 8 \ln (2) A (L_\mathrm{bs}/ \lambda_q)^2 e^{-b^2}
\left\{    
\sqrt{\pi}     \Re{ \left[ (l - i b) \erf{(l - i b)}\right] } + \sqrt{\pi} b \ \mathrm{erfi} {(b)} + e^{b^2} \left[ \cos{(2bl)} e^{-l^2} -1  \right]    
\right\} \, ,
   \label{tot_energy111}
\end{equation}
where
$$
A=\sqrt{32\pi} \Delta \omega (2 \pi \mathcal{F}_0)^2 \, .
$$

Using the Dawson function $F(b) \equiv \sqrt{\pi} \exp{-b^2} \mathrm{erfi}( b)  /2$, we rewrite Eq.~(\ref{tot_energy111}) as
\begin{equation}
W_q = 8 \ln (2) A (L_\mathrm{bs}/ \lambda_q)^2 
\left\{    
\sqrt{\pi} e^{-b^2}    \Re{ \left[ (l - i b) \erf{(l - i b)}\right] } + \cos{(2bl)} e^{-l^2} + 2 b  F(b) -1  \right\} \, .
   \label{tot_energy11}
\end{equation}

An asymptotic approximation of Eq.~\eqref{tot_energy11} can be found taking into account that 
\begin{equation}
1-\erf(y)\approx \frac{\exp{-y^2}}{ y \sqrt{\pi} }\left( 1-\frac{1}{2 y^2}\right)
\label{asymptotic}
\end{equation}
under $|y|^2 \gg 1$. For $y=l-ib$ the latter condition is written as 
\begin{equation}
    l^2 + b^2\gg 1 \, .
    \label{condition2}
\end{equation}
Using Eq.~\eqref{asymptotic}, we rewrite Eq.~\eqref{tot_energy11} as
\begin{equation}
W_q \approx 8 \ln(2) A\left( \frac{L_\mathrm{bs}}{\lambda_q}\right)^2
\left\{ \sqrt{\pi} l \exp(-b^2)-\left[ \frac{\cos(2 b l)}{2(b^2+4 l^2)}+ \frac{l \sin(2 b l)}{b^3+4 l^2}\right] \exp(-l^2) +2 b F(b)-1\right\} \, .
\label{tot_energy_long_L}
\end{equation}

\subsection{Generation at short propagation distances}
Expanding Eq.~\eqref{tot_energy11} in the Taylor series near $l=0$ up to $l^4$, we find
\begin{equation}
W_q \approx  A 8 \ln (2) (L_\mathrm{bs}/ \lambda_q)^2
\left\{  l^2  \left[1-\frac{l^2}{3}\left(b^2+\frac{1}{2}\right)\right]
\right\} \, .
\label{tot_energy2}
\end{equation}
The derivative over $l$ of the latter equation is zero under $l^2=3/(2b^2+1)$. Thus, for
\begin{equation}
 l^2  \ll \frac{3}{2 b^2+1}
 \label{condition1}
\end{equation}
one can keep only the first term in the square brackets in Eq.~\eqref{tot_energy2}, therefore we can write
\begin{equation}
W_q \approx 8 \ln (2) A (L_\mathrm{bs}/ \lambda_q)^2 l^2 
\label{tot_energy_short_L}
\end{equation}
 or
\begin{equation}
W_q \approx 
A \frac{\pi^2}{2} \left( \frac{L}{\lambda_q}\right)^2 \, .
\label{tot_energy_short_L1}
\end{equation}
Note that condition~\eqref{condition1} is satisfied for the small propagation distances and can be written as $$L \ll \min \left\{ L_\mathrm{coh}, L_\mathrm{bs} \right\}.$$

The Taylor series~\eqref{tot_energy2} can be used to describe $W_q$ for longer $l$. However, its accuracy decreases rapidly with $l$. Much better accuracy is achieved using Pad\'e approximation of Eq.~\eqref{tot_energy11} up to $l^2$
\begin{equation}
W_q \approx 8 \ln (2) A (L_\mathrm{bs}/ \lambda_q)^2 \frac{l^2}{1+(l^2/3)(b^2+1/2)} \, .
\label{Pade1}
\end{equation}

This approximation is used in Sec.~\ref{Results} to fit the numerical results up to distances $L \approx  L_\mathrm{bs}$. Omitting factors close to unity, the latter equation can be written as
\begin{equation}
W_q \approx A \frac{\pi^2}{2} \frac{L^2}{1+\frac{L^2}{3(2 L_\mathrm{coh}/\pi)^2}+ \frac{L^2}{6 L_\mathrm{bs}^2}} \, .
\label{Pade_approximation1}
\end{equation}

\subsection{Phase-matching defined generation}
Let us consider the case $b \gg 1$. From Eq.~\eqref{main_b} one can see that this condition corresponds to $L_\mathrm{coh} \ll L_\mathrm{bs}$, so the phase matching limits HHG efficiency. In this case Eq.~\eqref{tot_energy_long_L} is valid for all $l$ due to Eq.~\eqref{condition2}. Moreover, it can be further simplified. Namely, taking into account that for $b \gg 1$ the Dawson function can be approximated as $F(b) \approx 1/(2b)+1/(4 b^3)$, we have $2 b F (b) -1 \approx 1/(2 b^2)$. Thus, Eq.~\eqref{tot_energy_long_L} becomes
\begin{equation}
W_q \approx 8 \ln(2) A \left( \frac{L_\mathrm{bs}}{\lambda_q}\right)^2
\left\{ \sqrt{\pi} l \exp(-b^2)+\frac{1}{2b^2} - \left[ \frac{\cos(2 b l)}{2(b^2+4 l^2)}+ \frac{l \sin(2 b l)}{b^3+4 l^2}\right] \exp(-l^2) \right\} \, .
\label{tot_energy_long_L1}
\end{equation}

For $l \ll b$ we have
\begin{equation}
W_q \approx 8 \ln(2) A \left( \frac{L_\mathrm{bs}}{\lambda_q}\right)^2
\left\{ \frac{1}{2b^2} -  \frac{\cos(2 b l)}{2b^2} \exp(-l^2) \right\} 
\label{tot_energy_long_L11}
\end{equation}
or, substituting $l$ and $b$ from Eqs.~\eqref{l} and~\eqref{main_b}, we have
\begin{equation}
W_q \approx A  \left( \frac{L_\mathrm{coh}}{\lambda_q}\right)^2
\left\{ 1 - \cos\left(\frac{\pi L}{L_\mathrm{coh}}\right) \exp\left( - \frac{\pi^2 L^2}{16\ln(2) L_\mathrm{bs}^2}\right) \right\} 
\label{tot_energy_long_L2}
\end{equation}
(in agreement with Eq.~\eqref{tot_energy_short_L1} for short $L$).

For $l \gg b$ we have from Eq.~\eqref{tot_energy_long_L1}
\begin{equation}
W_q \approx  A\left( \frac{L_\mathrm{coh}}{\lambda_q}\right)^2
\left\{ 1 + 2 \sqrt{\pi} l b^2 \exp{-b^2} \right\} \, .
\label{tot_energy_long_L3}
\end{equation}
Note that under reasonable values of $l$ the first term dominates due to vanishing $\exp{-b^2}$, so $W_q$ almost saturates for long propagation distances. Thus, Eq.~\eqref{tot_energy_long_L2} can be used for all $L$.

\subsection{Blue-shift defined generation}
Let us consider the case 
\begin{equation}
      b \approx 1 \: \: \: \mathrm{or} \: \: \: b<1 \, .
\label {condition_bs}
\end{equation}
From Eq.~\eqref{main_b} one can see that this condition corresponds to $L_\mathrm{bs} \approx  L_\mathrm{coh}$ or $L_\mathrm{bs} <  L_\mathrm{coh}$, so the blue shift limits HHG efficiency. Under \eqref{condition_bs} the condition~\eqref{condition1} is written as $l^2 \ll 1$. $W_q$ is given by Eq.~\eqref{tot_energy_short_L1}. Under \eqref{condition_bs} the condition~\eqref{condition2} is written as $l^2 \gg 1$. From Eq.~\eqref{tot_energy_long_L} we have
\begin{equation}
W_q \approx 8 \ln(2) \sqrt{\pi} A\left( \frac{L_\mathrm{bs}}{\lambda_q}\right)^2 \, 
\left\{  l +d \right\}\exp{-b^2}. 
\label{tot_energy_long_L4}
\end{equation}
where
\begin{equation}
d=[2 b F(b)-1] \exp{b^2} / \sqrt{\pi}
\label{d}
\end{equation}
or
\begin{equation}
W_q \approx  2 \sqrt{\ln(2)} \pi^{3/2} A\left( \frac{L_\mathrm{bs}}{\lambda_q}\right) \left( \frac{L+D}{\lambda_q}\right) 
\exp{-\frac{4 \ln(2) L_\mathrm{bs}^2}{L_\mathrm{coh}^2}}
\, ,
\label{tot_energy_long_L5}
\end{equation}
where
\begin{equation}
D=\frac{4 \sqrt{\ln(2)} L_\mathrm{bs} [2 b F(b)-1] \exp{b^2}}{\pi^{3/2}} \, .
\label{D}
\end{equation}

Moreover, for $b < 1 $ using the Dawson function approximation (similar to~\cite{Dawson_approximation}) $F(b) \approx \frac{1-\exp(-b^2)}{2b}+\frac{b}{2}\exp(-b^2)$ we write Eq.~\eqref{d} as
\begin{equation}
d \approx  (b^2-1)/ \sqrt{\pi}
\label{d1}
\end{equation}
and
\begin{equation}
D \approx  \frac{4 \sqrt{\ln(2)} L_\mathrm{bs} [b^2-1]} {\pi^{3/2}} \, .
\label{D1}
\end{equation}

In a similar way approximate Eqs.~\eqref{int_short_L}, \eqref{int_all_L}, \eqref{int_long_L} for $I_q(L)$ are obtained from Eq.~\eqref{int_theory} using Taylor expansion of the error function near zero, or its asymptotic expansion~\eqref{asymptotic} for large arguments.

\section*{Acknowledgements}
This study was funded by RSF (grant No 22-22-00242). 
We acknowledge fruitful discussions with E.~Constant. 
We are grateful to V.~Birulia for the optimisation of the propagation code. 

\bibliography{lit}

\end{document}